\documentclass{article}
\usepackage{amsmath,graphicx,multirow,url,adjustbox,bm}
\usepackage{cite}
\usepackage{color,soul}
\usepackage{hyperref}
\usepackage{mathtools}
\usepackage{tabularx}
\usepackage{booktabs}
\usepackage{stfloats}
\usepackage[pdf]{graphviz}
\usepackage{pgfplots}
\pgfplotsset{width=\linewidth*1.04,compat=1.17} 

\newif\ifarxiv
\arxivtrue

\newcommand{\arxivGuard}[2]{
    \ifarxiv
        #1
    \else
        #2
    \fi
}

\arxivGuard{\usepackage[preprint]{spconf}}{\usepackage{spconf}}

\toappear{To appear in {\it Proc.\ ICASSP 2023, June 04-10, 2023, Rhodes island, Greece}}

\fnbelowfloat 

\usepackage{xpatch}
\makeatletter
\newcommand*{\addFileDependency}[1]{
  \typeout{(#1)}
  \@addtofilelist{#1}
  \IfFileExists{#1}{}{\typeout{No file #1.}}
}
\makeatother
\xpretocmd{\digraph}{\addFileDependency{#2.dot}}{}{}

\newcolumntype{Y}{>{\centering\arraybackslash}X}

\def\greps{\detokenize{〈}\detokenize{ε}\detokenize{〉}}
\def\grepsb{\detokenize{〈}\detokenize{ε}<SUB>b</SUB>\detokenize{〉}}
\def\grepse{\detokenize{〈}\detokenize{ε}<SUB>e</SUB>\detokenize{〉}}
\def\grblk{\detokenize{〈}b\detokenize{〉}}

\title{Powerful and Extensible WFST Framework for RNN-Transducer Losses}

\name{
  \vspace{-1pt} 
  \begin{tabular}{c} Aleksandr Laptev$^{1,2,*}$\thanks{$^*$Equal contribution.}, Vladimir Bataev$^{1,3,*}$, Igor Gitman$^1$, Boris Ginsburg$^1$
  \end{tabular}
}
\address{
  \vspace{-1pt} 
  $^1$NVIDIA, USA \\
  $^2$ITMO University, St. Petersburg, Russia \ \ \ \ \ \ \ \ \ \ \
  $^3$University of London, London, UK
  \vspace{-4pt} 
}
\begin{document}
%
\maketitle
\begin{abstract}
\vspace{-1pt} 
This paper presents a framework based on Weighted Finite-State Transducers (WFST) to simplify the development of modifications for RNN-Transducer (RNN-T) loss. Existing implementations of RNN-T use CUDA-related code, which is hard to extend and debug. WFSTs are easy to construct and extend, and allow debugging through visualization. We introduce two WFST-powered RNN-T implementations: (1) ``Compose-Transducer'', based on a composition of the WFST graphs from acoustic and textual schema -- computationally competitive and easy to modify; (2) ``Grid-Transducer'', which constructs the lattice directly for further computations -- most compact, and computationally efficient. We illustrate the ease of extensibility through introduction of a new W-Transducer loss -- the adaptation of the Connectionist Temporal Classification with Wild Cards. W-Transducer (W-RNNT) consistently outperforms the standard RNN-T in a weakly-supervised data setup with missing parts of transcriptions at the beginning and end of utterances. All RNN-T losses are implemented with the k2 framework and are available in the NeMo toolkit.
\end{abstract}

\vspace{-3pt} 
\vspace{-1pt} 
\vspace{-6pt} 
\section{Introduction}
\label{sec:intro}
\vspace{-4pt} 

RNN-Transducer (RNN-T) \cite{graves2012transducer}, along with Connectionist Temporal Classification (CTC) \cite{graves_connectionist_2006}, are the two most popular loss functions for end-to-end automatic speech recognition (ASR). Both objectives perform an alignment-free training by marginalizing over all possible alignments using a special $\langle blank \rangle$ unit. CTC allows repeated language units to indicate unit continuation in the time-frame while RNN-T does not, emitting exactly one non-$\langle blank \rangle$ prediction per unit. 

There are a large number of RNN-T variants: RNN-T with strictly monotonic alignment \cite{sak17_interspeech,tripathi19_asru,moritz2022sequence}, RNN-T optimized for better latency or compute \cite{yu21_icassp,mahadeokar21slt,tian21_interspeech,kim21j_interspeech,shinohara22_interspeech,jia22_interspeech,kuang22_interspeech}, or variants which improve accuracy for streaming or with external language model \cite{variani2022global,variani20_icassp,weng20_interspeech,guo2020_interspeech}, etc\footnote{\vspace{-1pt}A list of commonly used implementations of the RNN-T loss function \vspace{-1pt}can be found in the ``transducer-loss-benchmarking'' repository at\\
\begin{scriptsize}\url{https://github.com/csukuangfj/transducer-loss-benchmarking}\end{scriptsize}}. Most RNN-T loss implementations have thousands of lines. The modification of these implementations is hard and difficult to debug, since it usually requires the solid knowledge of CUDA or Numba \cite{lam2015numba}.

On the other side, CTC loss can be naturally represented with the Weighted Finite-State Transducers (WFST)\cite{miao15_eesen, xiang19_crf, povey_k2,hannun2020differentiable,moritz21_icassp, laptev22ctc}. In WFST paradigm, all possible alignments form a lattice on which forward and backward scores are calculated. The lattice can contain $\langle \epsilon \rangle$-arcs (virtual arcs without input labels for structural purposes) needed in some loss modifications \cite{laptev22ctc}. Frameworks like k2\footnote{\begin{scriptsize}\url{https://github.com/k2-fsa/k2}\end{scriptsize}} or GTN\footnote{\begin{scriptsize}\url{https://github.com/gtn-org/gtn}\end{scriptsize}} accelerate WFST operations and their automatic differentiation on GPU. They  enable easily modifiable CTC loss implementation in a few dozen lines of Python code, which are also  computationally efficient. They also allow ``visual debugging'' since each loss component can be plotted as a graph.

This work is an investigation on WFST representations of RNN-T loss. There are existing WFST-based implementations of restricted RNN-T variants. RNN-Transducers from \cite{moritz2022sequence} can emit maximum one non-$\langle blank \rangle$ unit per frame. The streaming-intended RNN-T from \cite{variani2022global} contradicts forward-backward derivations of \cite{graves2012transducer}. Unlike previous works, our WFST-powered RNN-T supports unrestricted, original Transducer from \cite{graves2012transducer}. We start our exploration with proposing ``Epsilon-Transducer'' which emulates RNN-T loss on a CTC pipeline with $\langle \epsilon \rangle$-arcs. Refining this paradigm, we propose two more efficient WFST-powered RNN-T implementations:
  \begin{enumerate}
    \vspace{-2pt} 
    \item ``Compose-Transducer'', which composes the RNN-T lattice from acoustic and textual schema WFSTs.
    \vspace{-4pt} 
    \item ``Grid-Transducer'', which creates the lattice directly.
    \vspace{-2pt} 
  \end{enumerate}
In all modifications, certain conventions for labels apply, since the neural network output is binded to graphs by these labels. All code associated with this binding is separate and independently tested and does not require any changes related to the loss customization itself. To demonstrate the ease of modification of WFST-based RNN-T, we implement a new loss, Transducer with Wild Cards (W-Transducer, or W-RNNT), inspired by W-CTC \cite{cai2022wctc}.

This work uses the k2 framework for the WFST operations and forward-backward score calculation on RNN-T representations. The code is released as a part of the NeMo toolkit\footnote{\begin{scriptsize}\url{https://github.com/NVIDIA/NeMo}\end{scriptsize}}~\cite{kuchaiev2019nemo}.

\begin{figure}[t!]
\arxivGuard{
    \centering\includegraphics[scale=0.4]{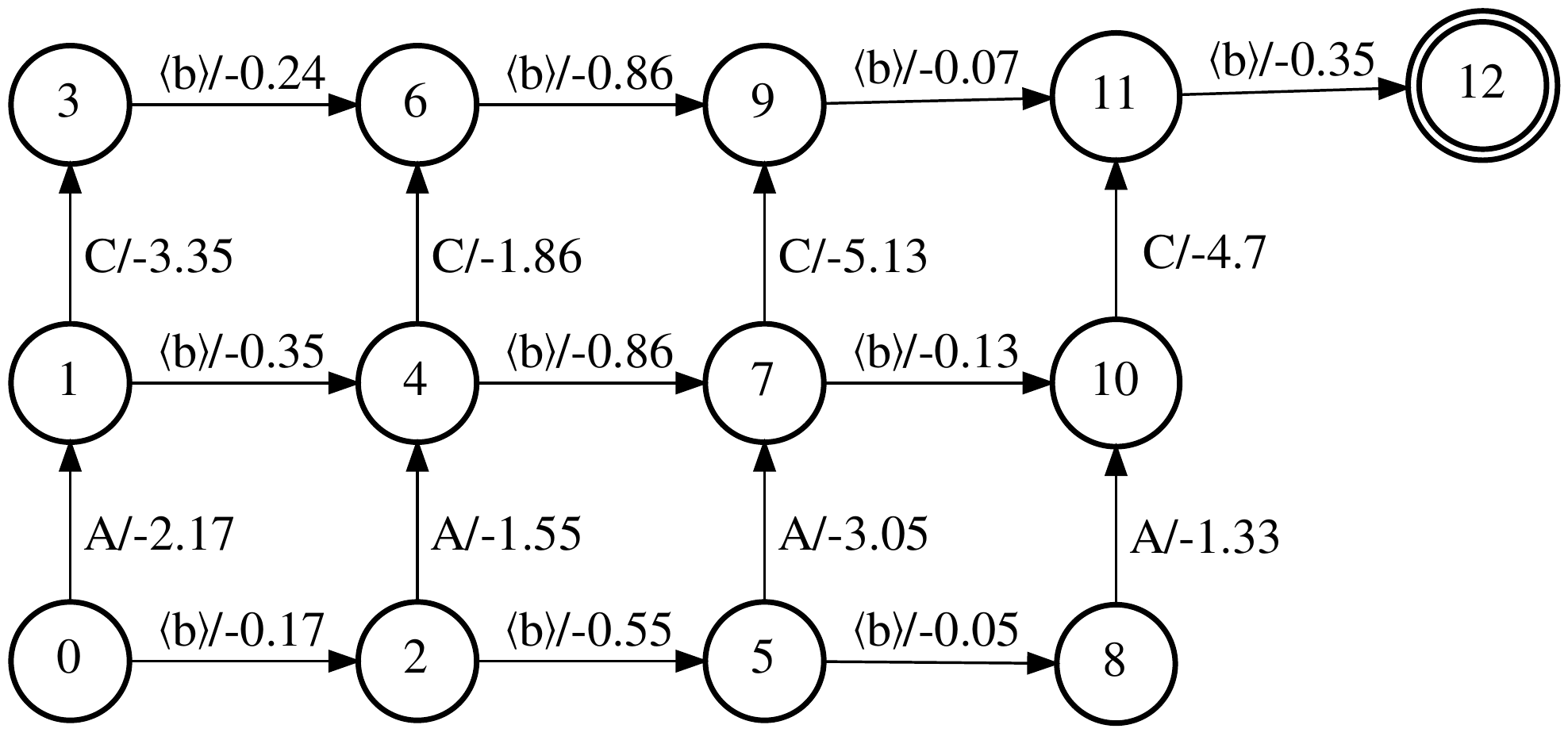}
}{
\centering\digraph[scale=0.4]{lattice32}{
    graph [fontname = "times" fontsize=18];
    node [fontname = "times" fontsize=18];
    edge [fontname = "times" fontsize=18];
	graph [center=1 nodesep=0.4 orientation=Portrait rankdir=LR ranksep=0.3 margin=0.0]
    node [height=0.6,
    width=0.6,
    shape=circle,
    style=bold];
    {
        rank=same;
		0	[label=3];
		1	[label=1];
		2	[label=0];
	}
	{
		rank = same;
		3	[label=6];
		4	[label=4];
		5	[label=2];
	}
	{
		rank = same;
		6	[label=9];
		7	[label=7];
		8	[label=5];
	}
	{
		rank = same;
		9	[label=11];
		10	[label=10];
		11	[label=8];
	}
	{
		rank = same;
		12	[label=12 shape=doublecircle];
		13	[label=13 style=invis];
		14	[label=14 style=invis];
	}
    0 -> 1	[headlabel="C/-3.35" dir=back labeldistance=4 labelangle=-55];
	1 -> 2	[headlabel="A/-2.17" dir=back labeldistance=4 labelangle=-55];
	0 -> 3	[label="\grblk/-0.24"];
	3 -> 4	[headlabel="C/-1.86" dir=back labeldistance=4 labelangle=-55];
	3 -> 6	[label="\grblk/-0.86"];
	1 -> 4	[label="\grblk/-0.35"];
	2 -> 5	[label="\grblk/-0.17"];
	4 -> 5	[headlabel="A/-1.55" dir=back labeldistance=4 labelangle=-55];
	4 -> 7	[label="\grblk/-0.86"];
	5 -> 8	[label="\grblk/-0.55"];
	6 -> 7	[headlabel="C/-5.13" dir=back labeldistance=4 labelangle=-55];
	6 -> 9	[label="\grblk/-0.07"];
	7 -> 8	[headlabel="A/-3.05" dir=back labeldistance=4 labelangle=-55];
	7 -> 10	[label="\grblk/-0.13"];
	8 -> 11	[label="\grblk/-0.05"];
	9 -> 10	[headlabel="C/-4.7" dir=back labeldistance=4 labelangle=-55];
	9 -> 12	[label="\grblk/-0.35"];
	10 -> 11	[headlabel="A/-1.33" dir=back labeldistance=4 labelangle=-55];
	10 -> 13	[style=invis];
	11 -> 14	[style=invis];
	13 -> 14	[style=invis];
	12 -> 13	[style=invis];
}
}
\vspace{-11pt} 
\caption{\hspace{-2pt}RNN-T lattice $\mathcal{L}$ example for a four-frame utterance ``A C'' and vocabulary $\langle b \rangle$, $A$, $B$, and $C$. $\langle b \rangle$ states for $\langle blank \rangle$.}
\label{fig:lattice}
\vspace{-5pt} 
\end{figure}

\section{WFST-powered RNN-T implementations}
\label{sec:wfst}
\vspace{-4pt} 

Most of the Maximum Likelihood (ML) based loss functions for the alignment-free ASR task (e.g. CTC) can be represented in the WFST paradigm as a negative forward scores of the following WFST composition:
\vspace{-6pt} 
\begin{equation}
 \label{equation:ml_loss}
    \textit{Loss}_{ml}(\bm{X},Y) = -\textit{Fwd}\Bigl(\mathcal{E}_{ml}(\bm{X}) \circ \bigl(\mathcal{T}_{ml} \circ \mathcal{Y}({Y})\bigl)\Bigl),
\vspace{-6pt} 
\end{equation}
where $\mathcal{E}_{ml}(\bm{X})$ is an Emissions graph over the log-probabilities tensor $\bm{X}$, $\mathcal{T}_{ml}$ is a loss Topology, and $\mathcal{Y}({Y})$ is a linear graph of the target unit sequence (unit graph)\footnote{\vspace{-1pt}The proposed representation is different from \cite{variani2022global}, which consists of the \vspace{-1pt}context dependency FSA, the alignment lattice FSA, and the weight function.}. 
Both Emissions and Topology graphs are loss-specific. 

Let's also denote $\mathcal{T} \circ \mathcal{Y}$ as an Alignment graph $\mathcal{A}$, and $\mathcal{E} \circ \mathcal{A}$ as a training Lattice $\mathcal{L}$ (see Fig.~\ref{fig:lattice}). Examples of CTC graphs $\mathcal{E}$, $\mathcal{Y}$, and $\mathcal{A}$ can be found in \cite{hannun2020differentiable}, and $\mathcal{T}$ in \cite{laptev22ctc}.

\vspace{-1pt} 
\subsection{Epsilon-Transducer}
\vspace{-4pt} 

\begin{figure}[b!]
\vspace{-6pt} 
\arxivGuard{
    \centering\includegraphics[scale=0.5]{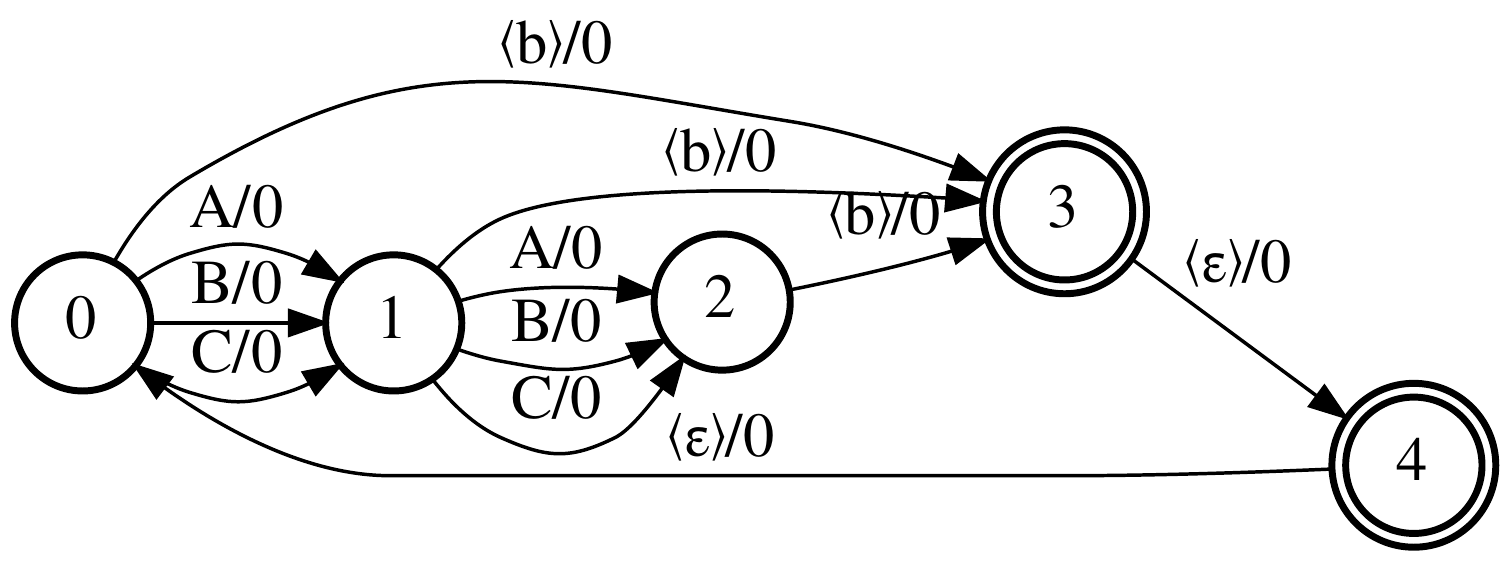}
}{
\centering\digraph[scale=0.5]{adapter2}{
    graph [fontname = "times" fontsize=18];
    node [fontname = "times" fontsize=18];
    edge [fontname = "times" fontsize=18];
	graph [center=1 nodesep=0.0 orientation=Portrait rankdir=LR ranksep=0.3 margin=0.0]
    node [height=0.5,
    width=0.5,
    shape=circle,
    style=bold];
    subgraph {
      0
    }
    3   [shape=doublecircle];
    4   [shape=doublecircle];
    0 -> 1  [label="A/0"];
    0 -> 1  [label="B/0"];
    0 -> 1  [label="C/0"];
    0 -> 3  [label="\grblk/0"];
    1 -> 2  [label="A/0"];
    1 -> 2  [label="B/0"];
    1 -> 2  [label="C/0"];
    1 -> 3  [label="\grblk/0"];
    2 -> 3  [label="\grblk/0"];
    3 -> 4  [label="\greps/0"];
    4 -> 0  [label="\greps/0"];
}
}
\vspace{-11pt} 
 \caption{Epsilon-Transducer Adapter: $\mathcal{D}^{\langle \epsilon \rangle}$ for a two-unit utterance and vocabulary size four: $\langle b \rangle$, $A$, $B$, and $C$.}
\label{fig:adapter}
\vspace{-3pt} 
\end{figure}

To make a WFST-powered RNN-T, one needs to define Emissions  $\mathcal{E}$ and Topology  $\mathcal{T}$ graphs. ``Minimal-CTC'' topology described in  \cite{laptev22ctc} fits RNN-T perfectly. Since there is no support for custom Emissions graphs in k2 and Kaldi \cite{povey11_kaldi}, we make RNN-T output log-probabilities tensors to fit the $\mathcal{E}$ abstraction in every framework. First, we squeeze audio and text dimensions $T$ and $U$ into one $T*U$. Then we emulate $\langle blank \rangle$ behavior (to jump over $U$ to the next time-frame) with $\langle \epsilon \rangle$-transitions. The Alignment graph will be as follows:
\vspace{-6pt} 
\begin{equation}
\label{equation:alignment_adapter}
    \mathcal{A}^{\langle \epsilon \rangle}_{rnnt}(Y) = \mathcal{D}^{\langle \epsilon \rangle}({Y}) \circ \bigl(\mathcal{T}_{rnnt} \circ \mathcal{Y}({Y})\bigl),
\vspace{-3pt} 
\end{equation}
where $\mathcal{D}^{\langle \epsilon \rangle}({Y})$ is an RNN-T $\langle \epsilon \rangle$ emissions adapter (see Fig.~\ref{fig:adapter}). Adapters are specific to the $U$ length. Since $\langle \epsilon \rangle$-transitions of $\mathcal{E}_{rnnt}(\bm{X})$ are either virtual or attached with probability one, they do not affect the lattice scores and therefore the loss and gradients. This gives numerical equivalence to classical RNN-T for model training and forced alignment tasks. The Epsilon-Transducer can be easily customized, but it it slow since the size of $\mathcal{A}^{\langle \epsilon \rangle}$ grows quadratically with the length of $U$, which makes calculating $\mathcal{L}$ as $\mathcal{E} \circ \mathcal{A}$ expensive.

\vspace{-1pt} 
\subsection{Compose-Transducer}
\vspace{-4pt} 

\begin{figure}[t]
\vspace{-2pt} 
\minipage[t]{\linewidth}
\arxivGuard{
    \centering\includegraphics[scale=0.5]{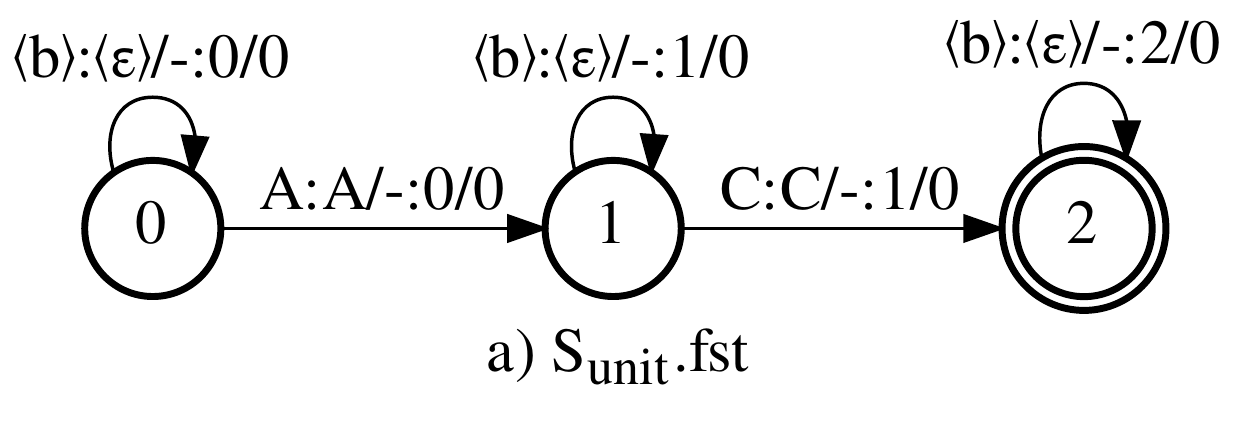}
}{
\centering\digraph[scale=0.5]{unitschema2}{
    graph [fontname = "times" fontsize=18];
    node [fontname = "times" fontsize=18];
    edge [fontname = "times" fontsize=18];
	graph [label=<a) S<SUB>unit</SUB>.fst> center=1 nodesep=0.25 orientation=Portrait rankdir=LR ranksep=0.3 margin=0.0]
    node [height=0.5,
    width=0.5,
    shape=circle,
    style=bold];
	0 -> 0	[label="\grblk:\greps/-:0/0"];
	0 -> 1	[label="A:A/-:0/0"];
	1 -> 1	[label="\grblk:\greps/-:1/0"];
	2	[shape=doublecircle];
	1 -> 2	[label="C:C/-:1/0"];
	2 -> 2	[label="\grblk:\greps/-:2/0"];
}
}
\label{fig:schema_unit}
\endminipage\hfill
\vspace{-13pt} 
\minipage[t]{\linewidth}
\arxivGuard{
    \centering\includegraphics[scale=0.4]{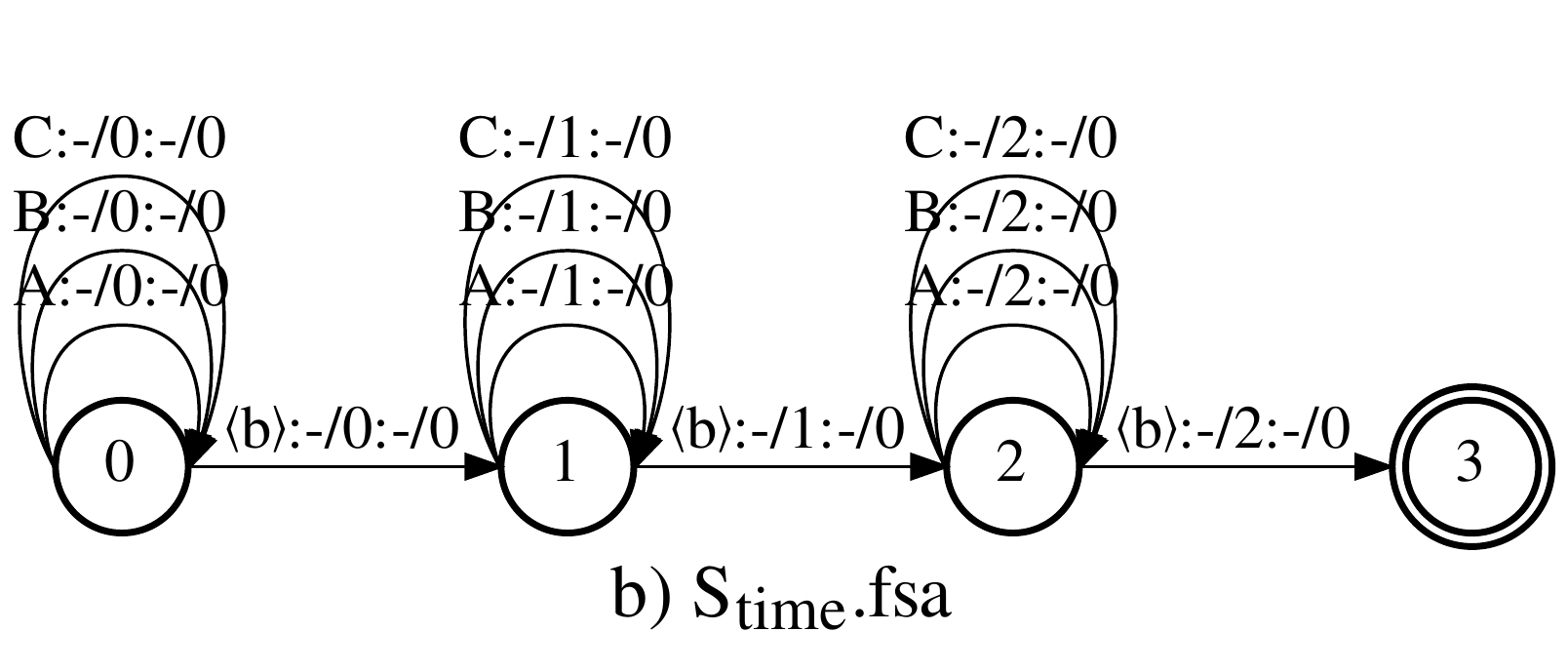}
}{
\centering\digraph[scale=0.4]{temporalschema3}{
    graph [fontname = "times" fontsize=22];
    node [fontname = "times" fontsize=18];
    edge [fontname = "times" fontsize=18];
	graph [label=<b) S<SUB>time</SUB>.fsa> center=1 nodesep=0.3 orientation=Portrait rankdir=LR ranksep=0.3 margin=0.0]
    node [height=0.5,
    width=0.5,
    shape=circle,
    style=bold];
	0 -> 0	[label="A:-/0:-/0" headport=e tailport=w];
	0 -> 0	[label="B:-/0:-/0" headport=e tailport=w];
	0 -> 0	[label="C:-/0:-/0" headport=e tailport=w];
	0 -> 1	[label="\grblk:-/0:-/0"];
	1 -> 1	[label="A:-/1:-/0" headport=e tailport=w];
	1 -> 1	[label="B:-/1:-/0" headport=e tailport=w];
	1 -> 1	[label="C:-/1:-/0" headport=e tailport=w];
	1 -> 2	[label="\grblk:-/1:-/0"];
	2 -> 2	[label="A:-/2:-/0" headport=e tailport=w];
	2 -> 2	[label="B:-/2:-/0" headport=e tailport=w];
	2 -> 2	[label="C:-/2:-/0" headport=e tailport=w];
	2 -> 3	[label="\grblk:-/2:-/0"];
	3 [shape=doublecircle];
}
}
\label{fig:schema_time}
\endminipage\hfill
\vspace{-14pt} 
\caption{Compose-Transducer: lattice schemas for a three-frame utterance ``A C'' with vocabulary $\langle b \rangle$, $A$, $B$, and $C$. a) Unit schema $\mathcal{S}_{unit}$, b) Time schema $\mathcal{S}_{time}$. Arc labels:\\
\texttt{$input$ : $output$ / $time$ : $unit\_idx$ / $weight$}.\\
``-'' indicates omitted labels.}
\label{fig:schemas}
\vspace{-5pt} 
\end{figure}

Calculating the RNN-T lattice $\mathcal{L}$ as $\mathcal{E} \circ \mathcal{A}$ only removes unnecessary arcs from $\mathcal{E}$. To remove the requirement for an explicit $\mathcal{E}$, we reformulate the lattice calculation as follows:
\vspace{-4pt} 
\begin{equation}
\label{equation:fast_rnnt}
    \mathcal{L}_{rnnt}(\bm{X},Y) = \textit{Populate}\Bigl(\mathcal{S}_{time}(\bm{X}) \circ \mathcal{S}_{unit}(Y),\bm{X}\Bigl),
\vspace{-4pt} 
\end{equation}
where $\mathcal{S}_{time}(\bm{X})$ is a temporal RNN-T Scheme and $\mathcal{S}_{unit}(Y)$ is a target unit sequence RNN-T Scheme (see Fig.~\ref{fig:schemas}). These schemas have additional labels besides input and output labels: $\mathcal{S}_{time}$ has time-frame indices and $\mathcal{S}_{unit}$ has unit indices (note that $\mathcal{S}_{unit}$ is just an $\mathcal{A}_{rnnt}$ with added unit indices). k2 can map these labels from composition components to the resulting WFST, which allows those indices (along with input label numbers) to be used to populate the arcs of $\mathcal{S}_{time} \circ \mathcal{S}_{unit}$ with the corresponding log-probabilities of $\bm{X}$ (here the $\textit{Populate}\Bigl(\mathcal{L}, \bm{X}\Bigl)$ method assigns scores to the lattice via indexed selection). The Compose-Transducer variant is almost as modular and customizable as the Epsilon-Transducer, but much faster and significantly smaller.

\vspace{-1pt} 
\subsection{Grid-Transducer}
\vspace{-4pt} 

The RNN-T lattice is a rectangular grid with temporal and target unit sequence axes. If the temporal axis is horizontal and the unit axis is vertical, then every horizontal arc is $\langle blank \rangle$ and each column of arcs is the same. Such a regular structure allows for direct lattice construction instead of the schemas composition. The Grid-Transducer is the most efficient and it remains a few dozen lines long. On the other hand, all the code related to graph construction must be implemented in vectorized operations to achieve high performance.

\section{W-Transducer}
\label{sec:w-transducer}
\vspace{-6pt} 

\begin{figure}[t!]
\vspace{-3pt} 
\arxivGuard{
    \centering\includegraphics[scale=0.4]{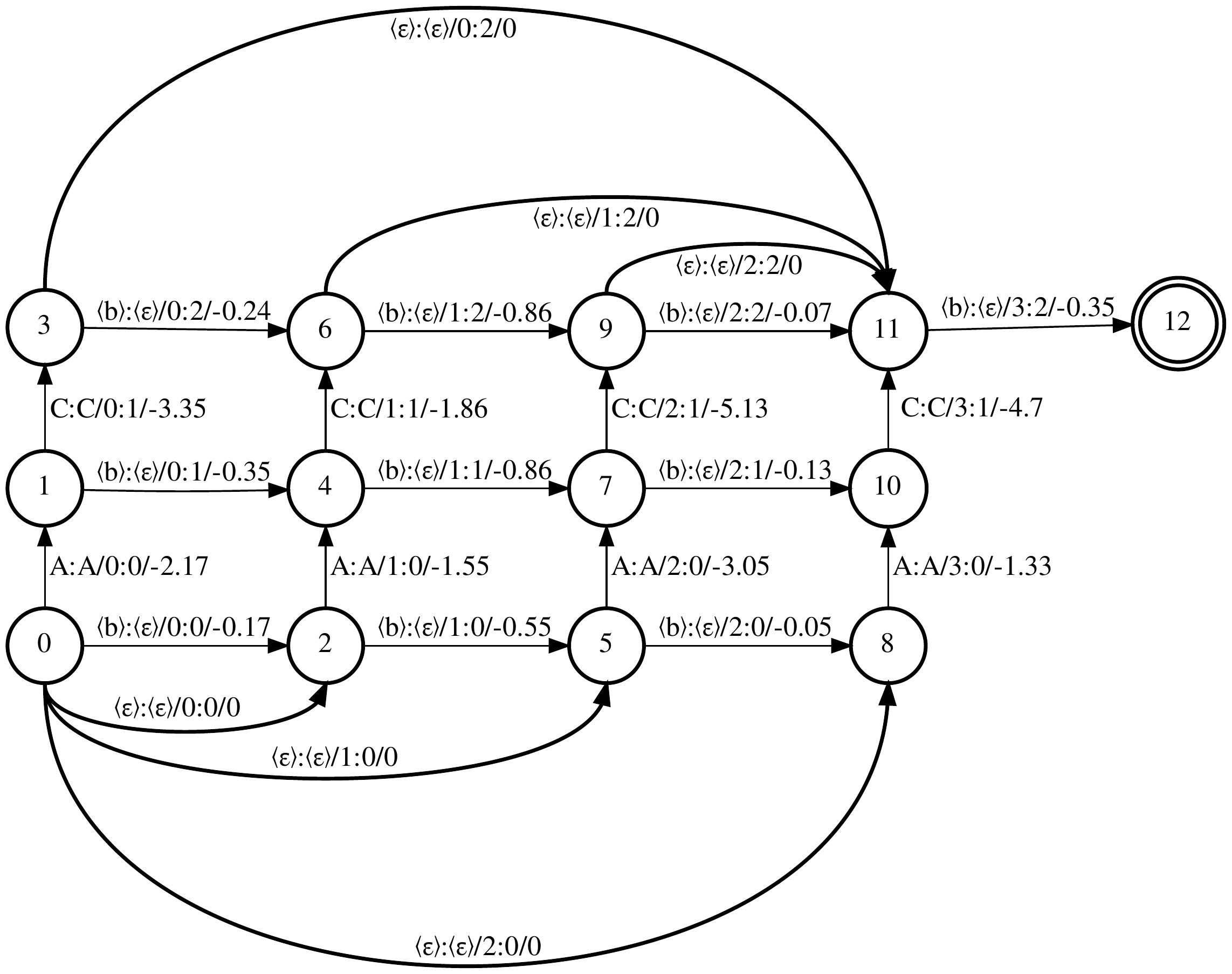}
}{
\centering\digraph[scale=0.4]{lattice32w}{
    graph [fontname = "times" fontsize=18];
    node [fontname = "times" fontsize=15];
    edge [fontname = "times" fontsize=15];
	graph [center=1 nodesep=0.3 orientation=Portrait rankdir=LR ranksep=0.0 margin=0.0]
    node [height=0.55,
    width=0.55,
    shape=circle,
    style=bold];
    {
        rank=same;
		0	[label=3];
		1	[label=1];
		2	[label=0];
	}
	{
		rank = same;
		3	[label=6];
		4	[label=4];
		5	[label=2];
	}
	{
		rank = same;
		6	[label=9];
		7	[label=7];
		8	[label=5];
	}
	{
		rank = same;
		9	[label=11];
		10	[label=10];
		11	[label=8];
	}
	{
		rank = same;
		12	[label=12 shape=doublecircle];
		13	[label=13 style=invis];
		14	[label=14 style=invis];
	}
    0 -> 1	[headlabel="C:C/0:1/-3.35" dir=back labeldistance=5 labelangle=-65];
	1 -> 2	[headlabel="A:A/0:0/-2.17" dir=back labeldistance=5 labelangle=-65];
	0 -> 3	[label="\grblk:\greps/0:2/-0.24"];
	3 -> 4	[headlabel="C:C/1:1/-1.86" dir=back labeldistance=5 labelangle=-65];
	3 -> 6	[label="\grblk:\greps/1:2/-0.86"];
	1 -> 4	[label="\grblk:\greps/0:1/-0.35"];
	2 -> 5	[label="\grblk:\greps/0:0/-0.17"];
	4 -> 5	[headlabel="A:A/1:0/-1.55" dir=back labeldistance=5 labelangle=-65];
	4 -> 7	[label="\grblk:\greps/1:1/-0.86"];
	5 -> 8	[label="\grblk:\greps/1:0/-0.55"];
	6 -> 7	[headlabel="C:C/2:1/-5.13" dir=back labeldistance=5 labelangle=-65];
	6 -> 9	[label="\grblk:\greps/2:2/-0.07"];
	7 -> 8	[headlabel="A:A/2:0/-3.05" dir=back labeldistance=5 labelangle=-65];
	7 -> 10	[label="\grblk:\greps/2:1/-0.13"];
	8 -> 11	[label="\grblk:\greps/2:0/-0.05"];
	9 -> 10	[headlabel="C:C/3:1/-4.7" dir=back labeldistance=5 labelangle=-65];
	9 -> 12	[label="\grblk:\greps/3:2/-0.35"];
	10 -> 11	[headlabel="A:A/3:0/-1.33" dir=back labeldistance=5 labelangle=-65];
	10 -> 13	[style=invis];
	11 -> 14	[style=invis];
	13 -> 14	[style=invis];
	12 -> 13	[style=invis];
	edge[weight=0];
	0 -> 3	[label=<\greps/0> style=invis headport=sw tailport=se];
	2 -> 5	[label=<\greps/0> style=invis headport=sw tailport=se];
	5 -> 8	[label=<\greps/0> style=invis headport=sw tailport=se];
	2 -> 5	[headlabel=<\greps:\greps/0:0/0> labeldistance=8 labelangle=-55 style=bold headport=s tailport=s];
	2 -> 8	[headlabel=<\greps:\greps/1:0/0> labeldistance=15 labelangle=-61 style=bold headport=s tailport=s];
	2 -> 11	[headlabel=<\greps:\greps/2:0/0> labeldistance=26 labelangle=-55 style=bold headport=s tailport=s];
	0 -> 9	[headlabel=<\greps:\greps/0:2/0> labeldistance=27 labelangle=57 style=bold headport=n tailport=n];
	3 -> 9	[headlabel=<\greps:\greps/1:2/0> labeldistance=16 labelangle=63 style=bold headport=n tailport=n];
	6 -> 9	[headlabel=<\greps:\greps/2:2/0> labeldistance=8 labelangle=56 style=bold headport=n tailport=n];
}
}
\vspace{-20pt} 
\caption{W-Transducer lattice $\mathcal{L}$ with time- and unit indices. Bold arcs represent skip-frame connections for wild cards.}
\label{fig:w_lattice}
\vspace{-5pt} 
\end{figure}

\subsection{CTC with Wild Cards}
\label{sec:w-ctc}
\vspace{-4pt} 

In weakly supervised learning, target unit sequence is incomplete, leading to a partial alignment problem: mapping the audio sequence to a unit sequence in which the target sequence is a sub-sequence. CTC with Wild Cards (W-CTC) \cite{cai2022wctc} partially solves this problem by training the model on a data with untranscribed beginnings and ends of the utterance. W-CTC introduces an optional virtual symbol ``*'', or the wild card, at the beginning of the sequence. Its per-frame probability is the sum of all unit probabilities, so the loss does not penalize the model for any inference. This allows for skipping (not forcing $\langle blank \rangle$ emissions) as many starting audio frames as needed. The forward-backward algorithm is also allowed to terminate immediately after aligning the partial target sequence and thus not force $\langle blank \rangle$ emissions in the remaining frames.
 
\vspace{-1pt} 
\subsection{W-Transducer}
\label{sec:w-transducer-impl}
\vspace{-4pt} 

W-CTC idea can be naturally adapted to the RNN-T loss\footnote{\vspace{-1pt}Recently proposed Star Temporal Classification (STC) \cite{pratap2022star} allows for \vspace{-1pt}labels missing at arbitrary positions, but its RNN-T adaptation would require \vspace{-1pt}modifying autoregressive Predictor besides the loss function.}. Fig.~\ref{fig:w_lattice} shows the lattice of one of the possible W-Transducer variants. Instead of sequentially iterating over wild card frames, we allow the algorithm to begin from any frame by introducing special skip-frame connections: $\langle \epsilon \rangle$ arcs with probability one from the initial state to any state before non-$\langle blank \rangle$ emissions.

The ability to terminate immediately after the partial alignment can also be implemented with skip-frame connections. The main difficulty here is that the RNN-T model must emit at least one $\langle blank \rangle$ unit after it finished emitting the last language unit. To ensure this rule, the ``final'' skip-frame connections, introduced after language units emission, point to the previous-to-final state.

\begin{figure}[t!]
\vspace{-9pt} 
\minipage[t]{\linewidth}
\arxivGuard{
    \centering\includegraphics[scale=0.5]{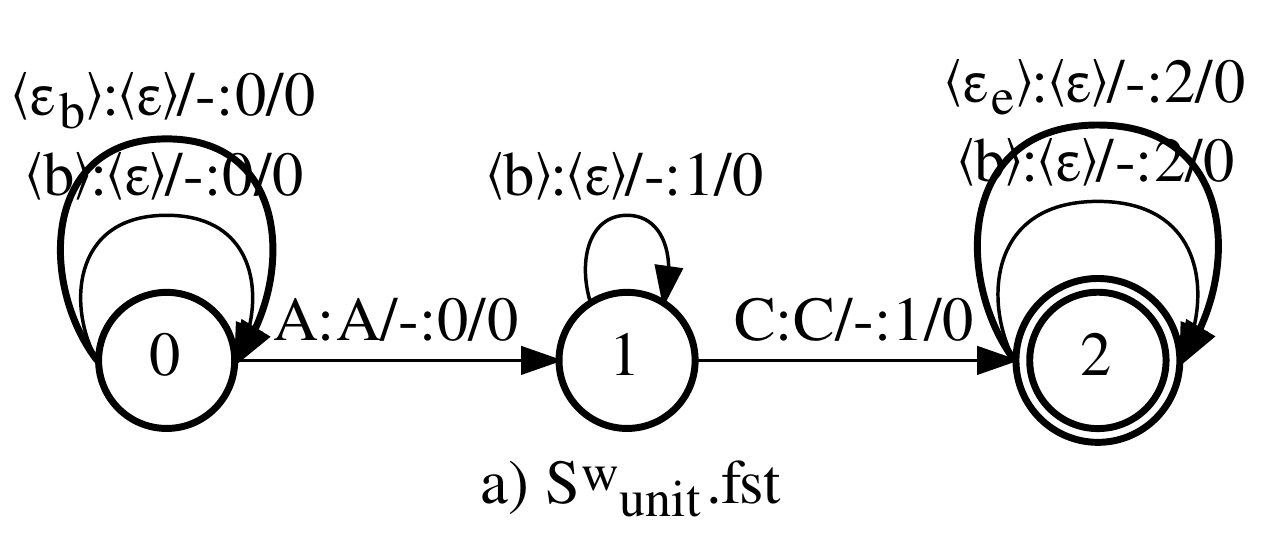}
}{
\centering\digraph[scale=0.5]{unitschema2w}{
    graph [fontname = "times" fontsize=18];
    node [fontname = "times" fontsize=18];
    edge [fontname = "times" fontsize=18];
	graph [label=<a) S<SUP>w</SUP><SUB>unit</SUB>.fst> center=1 nodesep=0.3 orientation=Portrait rankdir=LR ranksep=0.3 margin=0.0]
    node [height=0.5,
    width=0.5,
    shape=circle,
    style=bold];
	0 -> 0	[label="\grblk:\greps/-:0/0" headport=e tailport=w];
	0 -> 0	[label=<\grepsb:\greps/-:0/0> headport=e tailport=w style=bold];
	0 -> 1	[label="A:A/-:0/0"];
	1 -> 1	[label="\grblk:\greps/-:1/0"];
	2	[shape=doublecircle];
	1 -> 2	[label="C:C/-:1/0"];
	2 -> 2	[label="\grblk:\greps/-:2/0" headport=e tailport=w];
	2 -> 2	[label=<\grepse:\greps/-:2/0> headport=e tailport=w style=bold];
}
}
\label{fig:w_schema_unit}
\endminipage\hfill
\vspace{-2pt} 
\minipage[t]{\linewidth}
\arxivGuard{
    \centering\includegraphics[scale=0.4]{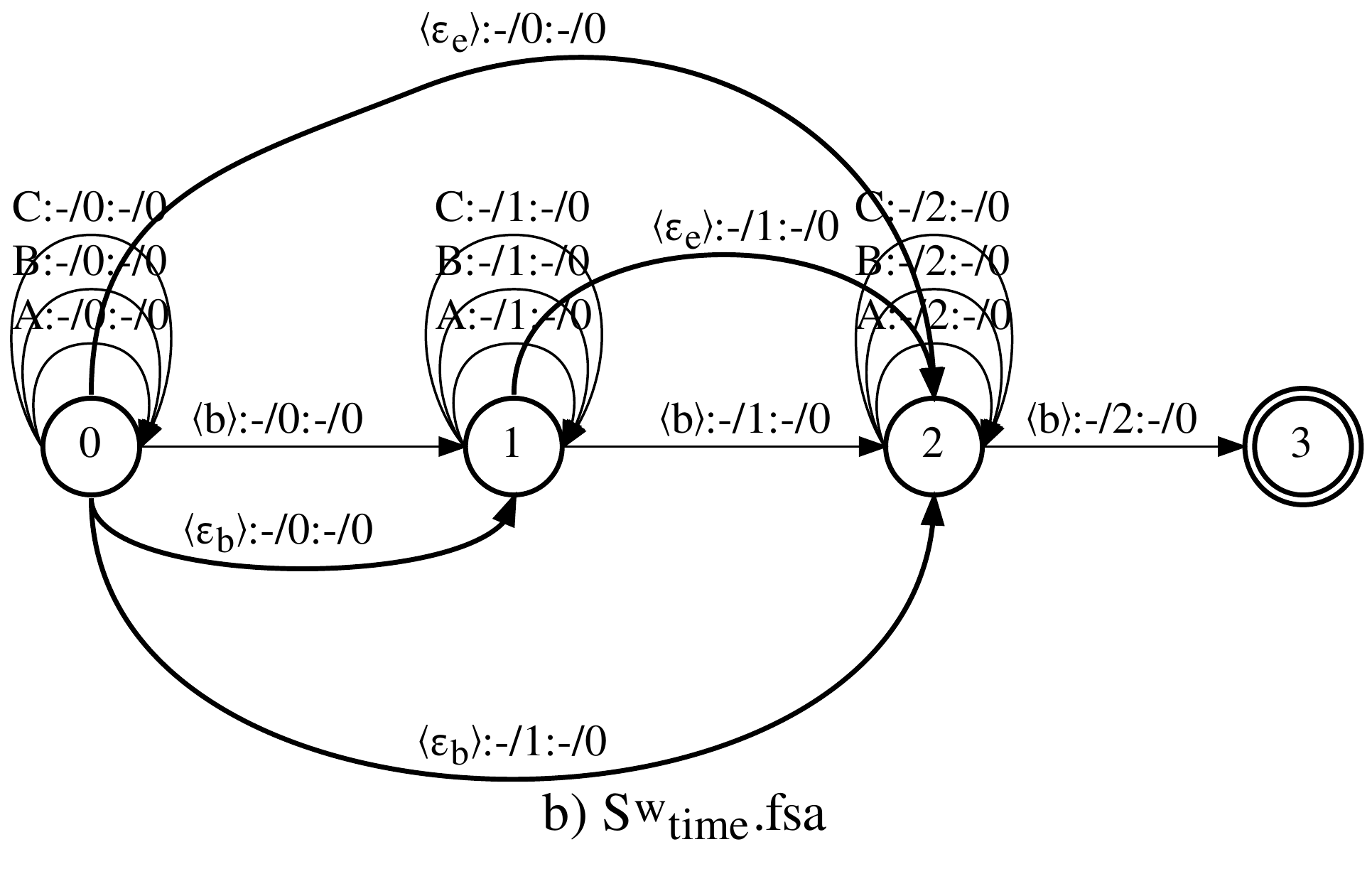}
}{
\centering\digraph[scale=0.4]{temporalschema3w}{
    graph [fontname = "times" fontsize=22];
    node [fontname = "times" fontsize=18];
    edge [fontname = "times" fontsize=18];
	graph [label=<b) S<SUP>w</SUP><SUB>time</SUB>.fsa> center=1 nodesep=0.3 orientation=Portrait rankdir=LR ranksep=0.5 margin=0.0]
    node [height=0.5,
    width=0.5,
    shape=circle,
    style=bold];
	0 -> 0	[label="A:-/0:-/0" headport=e tailport=w];
	0 -> 0	[label="B:-/0:-/0" headport=e tailport=w];
	0 -> 0	[label="C:-/0:-/0" headport=e tailport=w];
	0 -> 1	[label="\grblk:-/0:-/0"];
	1 -> 1	[label="A:-/1:-/0" headport=e tailport=w];
	1 -> 1	[label="B:-/1:-/0" headport=e tailport=w];
	1 -> 1	[label="C:-/1:-/0" headport=e tailport=w];
	1 -> 2	[label="\grblk:-/1:-/0"];
	2 -> 2	[label="A:-/2:-/0" headport=e tailport=w];
	2 -> 2	[label="B:-/2:-/0" headport=e tailport=w];
	2 -> 2	[label="C:-/2:-/0" headport=e tailport=w];
	2 -> 3	[label="\grblk:-/2:-/0"];
	3 [shape=doublecircle];
	edge[weight=0];
	0 -> 1	[label=<\grepsb:-/0:-/0> style=bold headport=s tailport=s];
	0 -> 2	[label=<\grepsb:-/1:-/0> style=bold headport=s tailport=s];
	0 -> 2	[label=<\grepse:-/0:-/0> style=bold headport=n tailport=n];
	1 -> 2	[label=<\grepse:-/1:-/0> style=bold headport=n tailport=n];
}
}
\label{fig:w_schema_time}
\endminipage\hfill
\vspace{-13pt} 
\caption{W-Compose-Transducer: lattice schemas with the skip-frame connections. Bold arcs represent the wildcards.}
\label{fig:w_schemas}
\vspace{-7pt} 
\end{figure}

This W-Transducer variant can be easily implemented with Compose-Transducer (see Fig.~\ref{fig:w_schemas}). Two skip-frame connections are added to the $\mathcal{S}^{w}_{unit}$ with different input labels to distinguish initial connections from final ones at $\mathcal{S}^{w}_{time}$.

\vspace{-1pt} 
\section{Experiments}
\label{sec:exp}
\vspace{-6pt} 

\vspace{-1pt} 
\subsection{Experimental setup}
\vspace{-4pt} 

We used the k2 framework to implement RNN-T and W-Transducer loss functions. RNN-T loss performance tests were carried out according to the transducer-loss-benchmarking pipeline\footnote{E.g. \href{https://github.com/csukuangfj/transducer-loss-benchmarking/blob/master/benchmark_k2.py}{\begin{scriptsize}\texttt{.../transducer-loss-benchmarking/.../benchmark\_k2.py}\end{scriptsize}}} with Encoder and Predictor embedding sizes 512, vocabulary size 500, batch size 30, and time and unit dimensions from 101 and 433 to 73 and 92 respectively. The measurements were taken on an NVIDIA V100 GPU.

The experiments with  W-Transducer were carried out in the NeMo toolkit. We used Conformer-Medium \cite{Gulati2020} with default hyper-parameters\footnote{\begin{scriptsize}\url{https://github.com/NVIDIA/NeMo/blob/main/examples/asr/conf/conformer/conformer_transducer_bpe.yaml}\end{scriptsize}} and LayerNorm normalization \cite{ba2016layer} and 1024 wordpiece vocabulary, trained on the 3-way speed-perturbed LibriSpeech \cite{panayotov2015librispeech}. The model was trained with global batch size 2048 for 200 epochs. We discarded 20\% and 50\% of the words from each training utterance in a random left-right proportion to simulate a weakly supervised conditions.

\vspace{-1pt} 
\subsection{RNN-T Benchmark results}
\vspace{-4pt} 

\begin{table}[t!]
\begin{center}
\adjustbox{max width=\linewidth}{
\begin{tabular}{ l | c | c c }
 \multicolumn{1}{c|}{\textbf{Implementation}} & \textbf{Prec.} & \multicolumn{1}{c|}{\textbf{Time, ms}} & \textbf{Memory, GB} \\
 \toprule
 \midrule
 \multirow{2}{*}{Warp-Transducer} & \textit{fp32} & 286 & 18.6 \\
  & \textit{fp16} & \multicolumn{2}{c}{not supported} \\
 \cmidrule(lr){1-2}
 \multirow{2}{*}{Warp-RNNT-Numba} & \textit{fp32} & 308 & 18.6 \\
  & \textit{fp16} & \multicolumn{2}{c}{not supported} \\
 \midrule
 \multirow{2}{*}{Epsilon-Transducer\textbf{*}} & \textit{fp32} & 83366 & 27.8 \\
  & \textit{fp16} & \multicolumn{2}{c}{not supported} \\
 \cmidrule(lr){1-2}
 \multirow{2}{*}{Compose-Transducer} & \textit{fp32} & 514 & 18.6 \\
  & \textit{fp16} & 350 & \textbf{17.1} \\
 \cmidrule(lr){1-2}
 \multirow{2}{*}{Grid-Transducer} & \textit{fp32} & 390 & 18.6 \\
  & \textit{fp16} & \textbf{231} & \textbf{17.1} \\
\end{tabular}
}
\end{center}
\vspace{-15pt} 
\caption{\label{tab:behcnmark-unsorted} Transducer Loss Benchmarking: unsorted batch.\\
\textbf{*} indicates that the loss was computed iteratively over batch to fit into GPU memory.}
\end{table}

\begin{table}[t]
\begin{center}
\adjustbox{max width=\linewidth}{
\begin{tabular}{ l | c | c c }
 \multicolumn{1}{c|}{\textbf{Implementation}} & \textbf{Prec.} & \multicolumn{1}{c|}{\textbf{Time, ms}} & \textbf{Memory, GB} \\
 \toprule
 \midrule
 \multirow{2}{*}{Warp-Transducer} & \textit{fp32} & 228 & 12.8 \\
  & \textit{fp16} & \multicolumn{2}{c}{not supported} \\
 \cmidrule(lr){1-2}
 \multirow{2}{*}{Warp-RNNT-Numba} & \textit{fp32} & 240 & 12.8 \\
  & \textit{fp16} & \multicolumn{2}{c}{not supported} \\
 \midrule
 \multirow{2}{*}{Compose-Transducer} & \textit{fp32} & 474 & 12.8 \\
  & \textit{fp16} & 345 & \textbf{11.8} \\
 \cmidrule(lr){1-2}
 \multirow{2}{*}{Grid-Transducer} & \textit{fp32} & 300 & 12.8 \\
  & \textit{fp16} & \textbf{167} & \textbf{11.8} \\
\end{tabular}
}
\end{center}
\vspace{-15pt} 
\caption{\label{tab:behcnmark-sorted} Transducer Loss Benchmarking: sorted batch.\\
Epsilon-Transducer results are omitted.}
\vspace{-5pt} 
\end{table}

We compared Epsilon-, Compose-, and Grid-Transducer with the first public RNN-T implementation ``Warp-Transducer'' and written in Python and Numba ``Warp-RNNT-Numba''. Note that while k2 supports only full (``\textit{fp32}'') and double (``\textit{fp64}'') precision operations, Compose- and Grid-Transducer implementations can populate the lattice with a half (``\textit{fp16}'') precision tensor and cast it to \textit{fp32} or \textit{fp64} only for the forward-backward score calculation.

Tables~\ref{tab:behcnmark-sorted} and ~\ref{tab:behcnmark-unsorted} show that Compose- and Grid-Transducer are as memory efficient as the prior implementations in \textit{fp32}. Switching to \textit{fp16} brings a 1.4x to 1.8x speed-up, making Compose-Transducer computationally competitive and Grid-Transducer up to 30\% faster than the CUDA-based Warp-Transducer. Epsilon-Transducer is inefficient in both speed and GPU memory, so it can only be used for prototyping.

\vspace{-1pt} 
\subsection{W-Transducer results}
\vspace{-4pt} 

Table~\ref{tab:wildcard-transducer-ls} demonstrates experimental validation of the W-Transducer against the original RNN-T on different weakly supervised conditions (with zero drop as the fully supervised setup). Here ``W-RNNT-force-final'' is the proposed W-Transducer variant from Section~\ref{sec:w-transducer-impl} and ``W-RNNT-allow-ignore'' is a variant in which the “final” skip-frame connections point to the final state (thus allowing to ignore $\langle blank \rangle$ emission after finishing language units emission).

The original RNN-T trained models were not able to reconstruct the true unit sequence, gaining Word Error Rate (WER) nearly one-to-one with the word drop percent. Both W-Transducer variants retained most of their accuracy even with 50\% of discarded words, which is consistent with the results from \cite{cai2022wctc}. The W-RNNT-force-final is better than the W-RNNT-allow-ignore variant in all scenarios. Yet, W-Transducer trained models degraded slightly (0.8\% absolute WER on \textit{test-other}) when no word is missing. We also noticed that the Wild Card variants are prone to overfitting when some labels are missing: the best WER results were obtained at early epochs ($<$100), after which the quality dropped significantly. But when all the text is present, no overfitting is observed.

\begin{table}[t!]
\begin{center}
\setlength{\tabcolsep}{3pt}
\adjustbox{max width=\linewidth}{
\begin{tabular}{ l | l | c c c c}
 \multirow{2}{*}{\textbf{Drop}} & \multicolumn{1}{c|}{\multirow{2}{*}{\textbf{Loss function}}} & \multicolumn{2}{c|}{\textbf{dev}} & \multicolumn{2}{c}{\textbf{test}} \\
 & & \textbf{clean} & \multicolumn{1}{c|}{\textbf{other}} & \textbf{clean} & \textbf{other} \\
 \toprule
 \midrule
  & RNN-T & \textbf{2.7} & \textbf{6.6} & \textbf{2.9} & \textbf{6.6} \\
 0\%  & W-RNNT-allow-ignore & 3.0 & 7.2 & 3.3 & 7.5 \\
   & W-RNNT-force-final & 3.0 & 7.1 & 3.2 & 7.4 \\
 \midrule
  & RNN-T & 25.0 & 27.9 & 25.3 & 28.2 \\
 20\% & W-RNNT-allow-ignore & 4.0 & 9.1 & 4.2 & 9.4 \\
  & W-RNNT-force-final & \textbf{3.8} & \textbf{8.8} & \textbf{4.1} & \textbf{9.0} \\
 \midrule
  & RNN-T & 54.6 & 56.5 & 55.0 & 56.9 \\
 50\% & W-RNNT-allow-ignore & 4.4 & 10.3 & 4.5 & 10.6 \\
  & W-RNNT-force-final & \textbf{4.0} & \textbf{9.0} & \textbf{4.1} & \textbf{9.4} \\
\end{tabular}
}
\end{center}
\vspace{-15pt} 
\caption{\label{tab:wildcard-transducer-ls} Different weakly supervised conditions against original RNN-T and W-Transducer variants, WER [\%].}
\vspace{-5pt} 
\end{table}

\vspace{-2pt} 
\section{Conclusion}
\label{sec:conclusion}
\vspace{-4pt} 
This paper presented an extensible framework for implementing and customizing RNN-T loss with WFST apparatus. This approach is easy to debug, and doesn't require GPU-specific code modification. Heavily based on simple and modular ``Epsilon-Transducer'' implementation, two efficient approaches, which construct the training lattice, populate it with relevant log-probabilities, and run the CUDA-accelerated forward and backward scores calculation, were presented:
``Compose-Transducer'' uses a composition of acoustic and textual schema WFSTs for the lattice construction. This approach is easy to modify and competitive in efficiency with conventional implementations.
``Grid-Transducer'' constructs lattice directly. This approach requires complex vectorized code to construct the graph, but can outperform conventional implementations by more than 25\% of speed and 7\% of memory consumption in half-precision calculations.

To illustrate the usefulness of the WFST paradigm for developing ASR loss functions, we define the Wild Card Transducer (W-Transducer) -- an RNN-T variant for weakly supervised learning. We introduced so-called wildcard arcs which indicate missing transcriptions at the beginning and the end of utterance.
W-Transducer gives about 68\% and 83\% relative WER improvement against the standard RNN-T when 20\% and 50\% of transcriptions are missing, respectively.


\bibliographystyle{IEEEbib}
\bibliography{refs}

\end{document}